\documentclass[twocolumn,showpacs,preprintnumbers,amsmath,amssymb]{revtex4}
\usepackage{graphicx}% Include figure files
\usepackage{epsfig}
\usepackage{dcolumn}% Align table columns on decimal point
\usepackage{bm}% bold math
\def\n {\nonumber}

\def\R  {{\cal R}}
\def\N  {{\cal N}}
\def\J {$J/\psi$ }
\def\Js {$J/\psi$'s }
\def\j { J/\psi  }

\def\pd {pseudo-diffractive}
\def\ktf {$k_t$-factorization }
\def\ktfa {$k_t$-factorization approach }

\newcommand{\bkappa}{\mbox{\boldmath $\kappa$}}

\newcommand{\bq}{\mbox{\boldmath $q$}}

\newcommand{\bk}{\mbox{\boldmath $k$}}

\newcommand{\Jot}{{\cal{J}}}
\def\cpc#1#2#3  {{Computer\ Phys.\ Comm.\ }  {\bf#1}, #2 (#3)}
\def\err#1#2#3  {{\it Erratum }              {\bf#1}, #2 (#3)}
\def\epjc#1#2#3 {{Eur. Phys. J. C }          {\bf#1}, #2 (#3)}
\def\dum#1#2#3  {{~}                         {\bf#1}, #2 (#3)}
\def\ib#1#2#3   {{\it ibid. }                {\bf#1}, #2 (#3)}
\def\jcp#1#2#3  {{J.\ Comput.\ Phys.\ }      {\bf#1}, #2 (#3)}
\def\jetpl#1#2#3 {{\rm JETP Lett.}           {\bf#1}, #2 (#3)}
\def\jhep#1#2#3 {{JHEP }                     {\bf#1}, #2 (#3)}
\def\ijmp#1#2#3 {{Int.\ J.\ Mod.\ Phys.\ }   {\bf#1}, #2 (#3)}
\def\jpg#1#2#3  {{J.\ Phys.\ G }             {\bf#1}, #2 (#3)}
\def\mpl#1#2#3  {{Mod.\ Phys.\ Lett.\ }      {\bf#1}, #2 (#3)}
\def\mpla#1#2#3 {{Mod.\ Phys.\ Lett.\ A }    {\bf#1}, #2 (#3)}
\def\ncim#1#2#3 {{Nuovo Cimento }            {\bf#1}, #2 (#3)}
\def\np#1#2#3   {{Nucl.\ Phys.\ }            {\bf#1}, #2 (#3)}
\def\npb#1#2#3  {{Nucl.\ Phys.\ B}           {\bf#1}, #2 (#3)}
\def\pan#1#2#3  {{Phys.\ At.\ Nuclei }       {\bf#1}, #2 (#3)}
\def\plb#1#2#3  {{Phys.\ Lett.\ B }          {\bf#1}, #2 (#3)}
\def\prep#1#2#3 {{Phys.\ Rep.\ }             {\bf#1}, #2 (#3)}
\def\prd#1#2#3  {{Phys.\ Rev.\ D }           {\bf#1}, #2 (#3)}
\def\prl#1#2#3  {{Phys.\ Rev.\ Lett.\ }      {\bf#1}, #2 (#3)}
\def\ptp#1#2#3  {{Prog.\ Theor.\ Phys.\ }    {\bf#1}, #2 (#3)}
\def\ps#1#2#3   {{Physica Scripta }          {\bf#1}, #2 (#3)}
\def\rmp#1#2#3  {{Rev.\ Mod.\ Phys.\ }       {\bf#1}, #2 (#3)}
\def\rpp#1#2#3  {{Rep.\ Prog.\ Phys.\ }      {\bf#1}, #2 (#3)}
\def\sa#1#2#3   {{Sci. Acta}                 {\bf#1}, #2 (#3)}
\def\sjnp#1#2#3 {{Sov.\ J.\ Nucl.\ Phys.\ }  {\bf#1}, #2 (#3)}
\def\spj#1#2#3  {{Sov.\ Phys.\ JETP }        {\bf#1}, #2 (#3)}
\def\spjl#1#2#3 {{Sov.\ JETP Lett.\ }        {\bf#1}, #2 (#3)}
\def\spu#1#2#3  {{Sov.\ Phys.-Usp.\ }        {\bf#1}, #2 (#3)}
\def\yaf#1#2#3  {{Yad.\ Fiz.\ }              {\bf#1}, #2 (#3)}
\def\zp#1#2#3   {{Zeit.\ Phys.\ }            {\bf#1}, #2 (#3)}
\def\zpc#1#2#3  {{Z.\ Phys.\ C }             {\bf#1}, #2 (#3)}
\def\etal {{\it et al. }}
\begin{document}
\title{Interparticle correlations in the production of \J pairs
       in proton-proton collisions}
\author{S.\ P.\ Baranov}
\email{baranov@sci.lebedev.ru}
\affiliation{P.N. Lebedev Institute of Physics, 
              Lenin Avenue 53, 119991 Moscow, Russia}
\author{A.\ M.\ Snigirev}
\email{snigirev@lav01.sinp.msu.ru}
\affiliation{Skobeltsyn Institute of Nuclear Physics,
     Lomonosov Moscow State University, 119991 Moscow, Russia}
\author{N.\ P.\ Zotov}
\email{zotov@theory.sinp.msu.ru}
\affiliation{Skobeltsyn Institute of Nuclear Physics,
      Lomonosov Moscow State University, 119991 Moscow, Russia}
\author{A.\ Szczurek}
\email{antoni.szczurek@ifj.edu.pl}
\affiliation{Institute of Nuclear Physics PAN, PL-31-342 Cracow, Poland}
\affiliation{University of Rzesz\'{o}w, PL-35-959 Rzesz\'{o}w, Poland}
\author{W.\ Sch\"{a}fer}
\email{wolfgang.schafer@ifj.edu.pl}
\affiliation{Institute of Nuclear Physics PAN, PL-31-342 Cracow, Poland}

\date{\today}
\begin{abstract}
We focus on the problem of 
disentangling the single (SPS) and double (DPS) parton 
scattering modes in the production of \J pairs at the LHC conditions.
Our analysis is based on comparing the shapes of the differential cross 
sections and on studying their behavior under imposing kinematical cuts.
On the SPS side, we consider the leading-order  ${\cal O}(\alpha_s^4)$
contribution with radiative corrections (taken into account in the
framework of the \ktf approach) and the subleading ${\cal O}(\alpha_s^6)$ 
contribution from \pd ~gluon-gluon scattering represented by one gluon 
exchange and two gluon exchange mechanisms. 
We come to the conclusion that disentangling the SPS and DPS modes is 
rather difficult on the basis of azimuthal correlations, while the rapidity 
difference looks more promising, provided the acceptance of the
experimental detectors has enough rapidity coverage.

\end{abstract}
\maketitle

%----------------------------
\section{Introduction}
%----------------------------

Since it was first observed, charmonium production in hadronic collisions
has been a subject of considerable theoretical interest. Production rates 
and their dependence on the different kinematic variables provide important
tests for comparing theoretical models.

In the last years, the production of \J pairs has attracted a significant 
renewal attention in the context of searches for double parton scattering 
processes \cite{Bartalini2011jp}. 
A number of discussions has been stimulated by the recent
measurement \cite{LHCb} of the double \J production cross section at the 
LHCb experiment at CERN. Theoretical estimates based on both collinear 
\cite{Berezhnoy,Novoselov,Stirling1} and \ktf \cite{Zotov} approaches
show that the single (SPS) and double (DPS) parton scattering contributions 
are comparable in size and, taken together, can perfectly describe the 
measured cross section. 

To disentangle the SPS and DPS mechanisms one needs to clearly understand
the production kinematics. Naive expectations that the SPS mechanism should 
result in the back-to-back event configuration received no support from the 
later calculations. Including the initial state radiation effects 
(either in the form of $k_t$-dependent gluon distributions \cite{BaranovJJ} 
or by means of simulating the parton showers in a phenomenological way
\cite{Stirling1}) washes out the original azimuthal correlations, thus 
making the SPS and DPS samples very similar to each other.
%(see also \cite{mpi}). 
One cannot exclude, however, that the situation may change under imposing 
certain cuts on the \J transverse momenta. On the other hand, it has been 
suggested \cite{Stirling1,Stirling2} that the DPS production is characterized
by a much larger rapidity difference between the two \J mesons.
The dominance of the DPS contribution over SPS at large rapidity
difference was discussed recently also for $pp\to c\bar{c}c\bar{c}X$ 
reaction \cite{LMS2012,SS2012}.
The goal of the present study is to carefully examine the \J pair
production properties in the different kinematical domains paying attention
to the different contributing processes.
On the SPS side, we consider the leading-order ${\cal O}(\alpha_s^4)$
subprocess (with radiative corrections taken into account in the
framework of the \ktf approach) and the subleading ${\cal O}(\alpha_s^6)$
contribution from \pd ~gluon-gluon scattering represented by one-gluon
exchange and two-gluon exchange mechanisms; the latter mechanisms yet have 
never been discussed in the context of searches for DPS.
On the DPS side, we consider the prompt production of \J pairs including
the direct $g+g\to\j+g$ contribution and radiative $\chi_c$ decays.

%-----------------------------------------
\section{Theoretical framework}

%--------------------------------------
\subsection{SPS contributions}
%--------------------------------------

At the leading order, ${\cal O}(\alpha_s^4)$, the SPS subprocess 
$g+g\to\j+\j$ is represented by a set of 31 "box" diagrams, 
with some examples displayed in Fig. 1. 
Our approach is based on perturbative QCD, 
nonrelativistic bound state formalism \cite{Chang,Baier,Berger}, 
and the \ktf ansatz \cite{GLR83,Catani,Collins} in the parton model.
The advantage of using the \ktfa comes from the ease of including the
initial state radiation corrections that are efficiently taken into 
account in the form of the evolution of gluon densities. 
The calculation of this subprocess is identical to that described
in Ref. \cite{BaranovJJ}.

As usual, the production amplitudes contain spin and color projection
operators that guarantee the proper quantum numbers of the final state
mesons. Then, the \J formation probability is determined by the radial 
wave function at the origin of coordinate space $|\R(0)|^2$; the latter 
is known from the \J leptonic decay width \cite{PDG}.
Only the color singlet channels are taken into consideration in the 
present study since this approach was found to be fully sufficient
\cite{BLZ} to describe all of the known LHC data on \J production.

The evaluation of Feynman diagrams is straightforward and follows the
standard QCD rules, with one reservation: in accordance with the \ktf
prescription \cite{GLR83}, the initial gluon spin density matrix is
taken in the form
%\begin{equation}
$\overline{\epsilon_g^{\mu}\epsilon_g^{*\nu}}=k_T^\mu k_T^\nu/|k_T|^2,$
%\end{equation}
where $k_T$ is the component of the gluon momentum perpendicular to the
beam axis. In the collinear limit, when $k_T\to 0$, this expression 
converges to the ordinary 
$\overline{\epsilon_g^{\mu}\epsilon_g^{*\nu}}=-g^{\mu\nu}/2$,
while in the case of off-shell gluons it contains an admixture of
longitudinal polarization.
All algebraic manipulations with Feynman diagrams have been done using
the computer system {\sc form} \cite{FORM}.

We have carefully checked that our present results are consistent with 
earlier calculations known in the literature. The model based on the 
diagrams of Fig. 1 was first formulated in Refs. \cite{Hmery,Scott}. 
Later on, it was extended to considering the polarization effects 
\cite{BaranovJung,Qiao} and to including the color-octet contributions,
see \cite{Qiao,Ko} and references therein. As far as the color-singlet 
contribution is concerned, all these papers are fully identical to each 
other; the calculations are made in the collinear scheme and restricted 
to the ${\cal O}(\alpha_s^4)$ order. Using the \ktfa 
%as in Ref. \cite{BaranovJJ} 
we go beyond these approximations by including the initial state 
radiation corrections.
We have checked that in the collinear limit we perfectly reproduce the 
results of Refs. \cite{Hmery,Scott,BaranovJung,Qiao,Ko}.

The full {\sc fortran} code for the matrix element is available from
the authors on request. This process is also available in the hadron 
level Monte Carlo generator {\sc cascade} \cite{CASCADE}.
Numerical results shown in the next section have been obtained using 
the $A0$ gluon distribution from \cite{Jung}.

%\subsection{Pseudodiffractive contributions}

In addition to the above, we also consider the \pd ~gluon-gluon 
scattering subprocesses represented by the diagrams of Fig. 2. 
Despite the latter are of formally higher order in $\alpha_s$, 
they contribute to the events with large rapidity difference between 
the two \J mesons and in that region can take over the leading-order 
'box' subprocess. Our processes differ from the true diffraction
in the sense that there occurs color exchange, and so, the rapidity
interval between the two \J's may be filled up with lighter hadrons
(thus showing no gap in the overall hadron density).
Among the variety of higher-order contributions, the \pd
~subprocesses mentioned here are of our special interest as they 
potentially can mimic the DPS mechanism having very similar kinematics.

The evaluation of the one-gluon exchange diagrams
$g(k_1)+g(k_2)\to\j(p_1)+\j(p_2)+g(k_3)+g(k_4)$ is straightforward, but 
the number of diagrams is rather large. There are six possible gluon
permutations in the upper quark loop and six permutations in the lower 
loop. Besides that, we have to consider interchanges between the two 
initial or two final gluons: 
$g(k_1)\leftrightarrow g(k_2)$, $g(k_3)\leftrightarrow g(k_4)$, 
thus ending up with 144 possible combinations.
Note that the matrix element is free from infrared singularities. 
This is due to the specific property of the quark loop amplitude which 
vanishes when any of the three attached gluons becomes soft.
These calculations have also been performed in the \ktfa as described 
above.

The two gluon exchange mechanism 
has been previously considered in Ref. \cite{Kiselev}, where it was
reduced to the production of \J pairs in photon-photon collisions 
\cite{Ginzburg} by recalculating the appropriate color factor.
We basically follow the same way in our present analysis,
but use an updated gluon density \cite{MSTW2008}.

Let us concentrate on the elementary $g{+}g{\to}\j{+}\j$ subprocess first. 
Only 16 of the different possible Feynman diagrams survive in the 
high-energy limit, which seems to be a suitable approximation for the 
conditions in discussion. The corresponding amplitude can be cast into 
the impact-factor representation \cite{Ginzburg}:
\begin{eqnarray}
&\hspace*{-4.5cm}
 A(g_{\lambda_1}g_{\lambda_2}\to V_{\lambda_3}V_{\lambda_4};s,t)=& \n \\
&is\! \displaystyle{\int}\! d^2 \bkappa\,\frac%
{\displaystyle{\Jot(g_{\lambda_1}{\to}V_{\lambda_3};\bkappa,\bq)\,
               \Jot(g_{\lambda_2}{\to}V_{\lambda_4};-\bkappa,-\bq)}}%
{\displaystyle{[(\bkappa+\bq/2)^2 +\mu_G^2]
               [(\bkappa-\bq/2)^2 +\mu_G^2]}}\,,&
\label{eq:A_2g}
\end{eqnarray}
and the cross section reads
\begin{equation}
{d\sigma(gg{\to}VV;s)\over dt} = 
%{1 \over 16 \pi s^2}\,{\N_c \over 4}\sum_{\lambda_i} \Big|
{\N_c \over 64 \pi s^2}\,\sum_{\lambda_i} \Big|
A(g_{\lambda_1}g_{\lambda_2}{\to}V_{\lambda_3}V_{\lambda_4};s,t)\Big|^2\, .
\end{equation}

Here the subscripts $\lambda_i$ denote the helicities of the gluons $g$
and vector mesons $V$, and $\bq$ is the transverse momentum transfer, 
$t\approx{-}\bq^2$. The overall color structure of the reaction 
is described by the factor $\N_c=(N_c^2{-}4)^2/[16N_c^2(N_c^2{-}1)]$,
where $N_c=3$. %is the number of colors.
We kept explicit an effective gluon mass $\mu_G$ which is responsible for
soft QCD effects \cite{Nikolaev:1993ke} and plays the 
%\cite{Nikolaev:1993ke,Kopeliovich:2007pq}
role of regularization parameter. However, in the present study we can 
safely set it to zero, and the amplitude remains finite as the impact 
factors $\Jot$ vanish when $\bkappa\to\pm\bq/2$. See also the discussion
in \cite{gamgam}.

At small $t$, within the diffraction cone, the cross section
is dominated by the $s$-channel helicity conserving amplitude.
In this case, the explicit form of the impact factor is
\begin{equation}
\Jot(g_{\lambda}{\to}V_{\tau};\bkappa,\bq) = \delta_{\lambda,\tau}
\sqrt{4\pi\alpha_s^3}\,\displaystyle{\int}
\frac{\displaystyle{\psi(z,\bk)I(z,\bk,\bq)}}%
     {\displaystyle{z(1-z)(2\pi)^3}}\,dz d^2\bk\, ,
\end{equation}
where $\psi(z,\bk)$ is the light-cone wave function of the vector 
meson and $z$ is the light-cone momentum fraction carried by the heavy 
quark. Neglecting the intrinsic motion of the quarks we set 
$\psi(z,\bk)=C\;\delta(z-\frac{1}{2})\,\delta^{(2)}(\bk)$,
where the normalizing constant $C$ is adjusted to the \J leptonic
width and is related to the radial wave function at the origin as
$C^2 = 12\pi^5/(N_c^2 m_\psi^3)|\R(0)|^2$. 
Within the above approximation, we have
\begin{equation}
I(z,\bk,\bq)= \frac{m_\psi}{2}\Big[
{1\over\bkappa^2 +m_\psi^2/4}-{4\over\bq^2 +m_\psi^2}\Big].
\end{equation}
Including the quark intrinsic motion would decrease the amplitude; 
for a more detailed analysis see \cite{gamgam}. As it will become 
clear from the numerical results, the nonrelativistic approximation 
is sufficiently accurate for our purposes.

The cross section for the two-gluon exchange contribution to the 
$p+p\to\j+\j+X$ reaction (see Fig. 2) is calculated in the
collinear approximation with MSTW2008(NLO) gluon distribution
function \cite{MSTW2008} and the factorization scale
$\mu_f^2 = m_t^2$, where $m_t$ is the \J transverse mass.
The elementary $g+g\to\j+\j$ cross section can be easily 
calculated in the high-energy approximation similarly to how it was 
done for the $\gamma+\gamma\to\j+\j$ reaction \cite{gamgam}. 
The corresponding cross section is proportional to 
$\alpha_s^6(\mu_r^2)$, and therefore depends strongly on the choice of 
the renormalization scale. In the calculation presented here we take 
$\mu_r^2 = m_t^2$. 
In the high-energy approximation, the matrix element is merely 
a function of the transverse momentum $\bq$ of one of the \J's. This 
cannot be true at low subprocess energy, close to the \J\J threshold. 
Here one 
must take into account also the longitudinal momentum transfer. We 
therefore replace $\bq^2$ by the exact $\hat t$ or $\hat u$ for the $t$ 
and $u$ diagrams respectively. We neglect here the possible interference 
between the box diagram and the two-gluon exchange mechanism, which is 
formally of lower order than the square of the two-gluon amplitude. 
However, firstly, the addition of box and two-gluon exchange amplitudes 
is not warranted without the consistent evaluation of 
$\alpha_s$-corrections to the box. Secondly, it will become obvious from 
the numerical results, that the two-gluon mechanism is exceedingly small 
in the region of invariant masses dominated by the box mechanism.

%--------------------------------------
\subsection{DPS contributions}
%--------------------------------------

Under the hypothesis of having two independent hard partonic subprocesses
$A$ and $B$ in a single $pp$ collision, and under further assumption that
the longitudinal and transverse components of generalized parton 
distributions factorize from each other, the inclusive DPS cross section 
reads 
(for details see, e.g., the recent review \cite{Bartalini2011jp} 
with many references to prior works listed therein)
\begin{equation} \label{doubleAB}
\sigma^{\rm AB}_{\rm DPS} = \frac{\displaystyle{m}}{\displaystyle{2}} 
\frac{\displaystyle{\sigma^{ A}_{\rm SPS}\sigma^{ B}_{\rm SPS}}}%
{\displaystyle{\sigma_{\rm eff}}},
\end{equation}
%\qquad
\begin{equation}
\sigma_{\rm eff}=\Bigl[ \int d^2b\,\bigl( T({\bf b}) \bigr)^2 \Bigr]^{-1},
\end{equation} 
where $T({\bf b}) = \int f({\bf b_1}) f({\bf b_1{-}b})\,d^2b_1 $ is the 
overlap function that characterizes the transverse area occupied by the 
interacting partons, and $f({\bf b})$ is supposed to be a universal 
function of the impact parameter ${\bf b}$ for all kinds of partons with 
its normalization fixed as
\begin{eqnarray} 
\label{f}
\int f({\bf b_1}) f({\bf b_1{-}b})\,d^2b_1\,d^2b=\int T({\bf b})\,d^2b=1.
\end{eqnarray} 
The inclusive SPS cross sections $\sigma^{A}_{\rm SPS}$ and 
$\sigma^{B}_{\rm SPS}$ for the individual partonic subrocesses $A$ and 
$B$ can be calculated in a usual way using the single parton distribution 
functions. The symmetry factor $m$ equals to 1 for identical subprocesses 
and 2 for the differing ones. 

These simplifying factorization assumptions, though rather customary 
in the literature and quite convenient from the computational point of 
view, are not sufficiently justified and are currently under revision
%\cite{Ryskin2011kk,Blok2010ge,Diehl2011tt,Diehl2011yj,stir,flesburg,%
%Blok2011bu,Gaunt2012wv,Manohar,Ryskin2012qx}. 
%\cite{Bartalini2011jp,Ryskin2011kk,flesburg}.
\cite{Bartalini2011jp}.
Nevertheless, we restrict ourselves to this simple form (\ref{doubleAB}) 
regarding it as the first estimate for the DPS contribution.
The presence of correlation term in the two-parton distributions 
results in reduction \cite{Ryskin2011kk,flesburg,Snigirev2010tk} 
of the effective cross section $\sigma_{\rm eff}$ with the growth of the 
hard scale, while the dependence of $\sigma_{\rm eff}$ on the total energy 
at a fixed scale is rather weak \cite{flesburg}. Thus, in fact, we obtain 
the lower bound estimate for the contribution under consideration. 
The CDF \cite{cdf4jets,cdf} and D0 \cite{D0} measurements give 
$\sigma_{\rm eff}\simeq$ 15 mb, that constitutes roughly 20$\%$ of the 
total (elastic + inelastic) $p{\bar p}$ cross section at the Tevatron 
energy. We will use this value in our further analysis.

When calculating the inclusive SPS cross section $\sigma^{\j}_{\rm SPS}$
we take into account both the direct production channel $g{+}g\to\j{+}g$ 
and the production of $P$-wave states $g{+}g{\to}\chi_{cJ}$ followed by 
radiative transitions $\chi_{cJ}{\to}\j{+}\gamma$. Numerically these two
production mechanisms turn to be of approximately equal importance.
The calculation of the relevant Feynman diagrams is straightforward, but
is done in the \ktfa implying that the initial gluon spin density matrix 
is taken in the form 
$\overline{\epsilon_g^{\mu}\epsilon_g^{*\nu}}=k_T^\mu k_T^\nu/|k_T|^2$
(similarly to what we did for the "box" SPS subprocess). The computational 
technique is explained in every detail in Ref. \cite{BLZ}.

The formation probability of \J meson is determined by its radial wave 
function; the latter is extracted from the known leptonic decay width 
\cite{PDG} and is set to $|\R_{\psi}(0)|^2=0.8$ GeV$^3$. The formation 
probability of $\chi_{cJ}$ mesons is determined by the derivative of 
the radial wave function; the latter is taken from the potential model 
\cite{EicQui}: $|\R_{\chi}'(0)|^2=0.075$ GeV$^5$. 
The decay branchings are taken from the Particle Data Book \cite{PDG}:
$Br(\chi_{c1}{\to}J/\psi\gamma)=35\%$, $Br(\chi_{c2}{\to}J/\psi\gamma)=20\%$.
The decay angular distributions are generated in accordance with the 
calculated $\chi_{cJ}$ polarization properties under the assumption of 
electric dipole dominance \cite{Wise}.

%-------------------------------------------
\section{Results and discussion}
%-------------------------------------------

We start with discussing the role of kinematic restrictions on the
\J transverse momentum. Shown in Fig. 3 are the fractions of SPS 
events surviving after imposing cuts on $p_T(\psi)$. 
Dashed line corresponds to requiring $p_T(\psi){>}p_{T,min}$ for only 
one (arbitrarily chosen) \J meson with no restrictions on the other.
Were the two \Js produced independently, the probability of having 
$p_T(\psi){>}p_{T,min}$ for the both \Js simultaneously could be 
obtained by just squaring the single-cut probability (dash-dotted 
curve in Fig. 3). 
On the contrary, in the naive the back-to-back kinematics, a cut
applied to any of the two \Js would automatically mean the same
restriction on the other, thus making no effect on the overall 
probability (dashed curve).
The DPS production mode with cuts applied to both \J mesons
is represented by the dotted curve in Fig. 3. As one can see, this 
curve is rather close to that modeling the idealized independent 
SPS production.

The explicit calculation (solid curve) lies between the two idealistic
extreme cases related to the fully independent (dash-dotted curve) and 
fully back-to-back correlated (dashed curve) production of \J pairs. 
In the region $p_{T,min}<4$ GeV the solid and dash-dotted curves almost 
coincide, thus showing that the two \Js are nearly idependent.
With stronger cuts on $p_T(\psi)$, the curves diverge showing
that the production of \Js becomes correlated.

Another illustration of this property is given by the distributions
in the azimuthal angle difference $d\sigma(\psi\psi)/d\Delta\varphi$ 
exhibited in Fig. 4.
The distribution looks flat for the unrestricted phase space (the 
upper plot), but tends to concentrate around $\Delta\varphi\simeq\pi$ 
when the cuts on $p_T(\psi)$ become tighter (the middle and the 
lower plots in Fig. 4.)
In principle, one could get rid of the SPS contribution by imposing
cuts like $p_T(\psi)>6$ GeV, $\Delta\varphi<\pi/4$, but the DPS cross
section would then fall from tens of nanobarns to few picobarns.
We can conclude that the SPS and DPS modes are potentially
distinguishable at sufficiently high $p_T(\psi)$, but the production 
rates fall dramatically, and so, the practical discrimination of the
production mechanisms remains problematic.

Now we turn to rapidity correlations explained in Figs. 5 and 6.
In the case of independent production (the DPS mode), the
distribution over $\Delta y$ is rather flat (dash-dotted curve in
Fig. 5), while in the case of SPS 'box' contribution (dotted curve 
in Fig. 5) it is concentrated around $\Delta y\simeq 0$ and does not 
extend beyond the interval $|\Delta y|<2$.
The shape of the double-differential cross section 
$d\sigma/dy(\psi_1)dy(\psi_2)$ corresponding to the Leading-Order SPS 
contribution is presented in Fig. 6. This contribution forms a long 
diagonal 'ridge' in the $y(\psi_1)-y(\psi_2)$ plane.

In Fig. 5 we also show \pd ~contributions from the one- and two-gluon 
exchange processes of Fig. 2.
As it was expected, these processes lead to relatively
large $\Delta y$ and even show maxima at $\Delta y \simeq\pm 2.$
This corresponds to a typical situation with one \J moving forward and 
the other one moving backward, following the directions of the initial 
gluons. The minimum for the two-gluon exchange $g+g\to\j+\j$ 
subprocess is a consequence of the educated guess correction
of the high-energy formula at low energies as discussed above.

At the same time, the absolute size of the one-gluon exchange cross
section is found to be remarkably small. There are several reasons 
taking credit for this smallness. First, is the presence of two extra 
powers of $\alpha_s$. Second, is just the large typical rapidity 
difference that makes the invariant mass of the final state relatively 
large:
$M_{\psi\psi}(\Delta{y}{=}2)/M_{\psi\psi}(\Delta{y}{=}0)
\simeq\cosh(\Delta{y}/2).$
In turn, larger masses mean larger values of the probed $x$, 
and, accordingly, lower values of the gluon densities. 

The third and the most important reason lies in the color factors. 
The color amplitude of the first diagram in the first row of Fig. 1 
reads $tr\{T^a T^c T^c T^b\}=[(N_c^2-1)/(4N_c)]\delta^{ab}$
%$\frac{2}{3}\delta^{ab}$
(where $T$ stand for the $SU(3)$ generators).
After taking square and summing over the initial gluon colors 
$a$ and $b$ it gives $[\frac{2}{3}\delta^{ab}]^2 = 32/9.$
Similarly, for the diagrams in the second row we have
$[tr\{T^aT^cT^d\}f^{bcd}]^2=[\frac{1}{4}f^{acd}f^{bcd}]^2=
[\frac{N_c}{4}\delta^{ab}]^2=9/2$ and 
$[tr\{T^cT^d\}f^{ace}f^{bde}]^2=[\frac{1}{2}f^{ace}f^{bce}]^2=
[\frac{N_c}{2}\delta^{ab}]^2=9.$ 

For comparison, the color amplitude of the first diagram in Fig. 2 
reads $\frac{1}{4}d^{ace}\cdot\frac{1}{4}d^{bde},$ and the terms 
containing $\frac{1}{4}f^{ace}$ or $\frac{1}{4}f^{bde}$ disappear 
because of cancellations between the different diagrams. 
This yields after squaring 
\begin{eqnarray}
[\frac{1}{4}d^{ace}\frac{1}{4}d^{bde}]^2 &= &
 \frac{(N_c^2{-}1)(N_c^2{-}4)^2}{256\;N_c^2} =
 \frac{1}{256}\frac{200}{9}\simeq 0.1. \nonumber
\end{eqnarray}
%which is much less than what we have from the diagrams in Fig. 1.
The color interference term is even smaller (and negative):
\begin{eqnarray}
\!\!\frac{\displaystyle{[d^{ace}d^{bde}][d^{ade}d^{bce}]}}%
{\displaystyle{256}}
&\!\!=\!\!&\frac{(N_c^2{-}1)(N_c^2{-}4)(N_c^2{-}12)}{512\;N_c^2}
 =\frac{-20}{256\cdot 3}. \nonumber
\end{eqnarray}
Note that all the considered contributions are of the same order 
in $N_c$.

The same suppression factors apply to the two-gluon exchange as well, 
but there is yet another suppressing mechanism  specific for the 
one-gluon exchange process. It comes from the fact that the amplitude 
vanishes when any of the final state gluons becomes soft 
(this property makes the process infrared-safe, as we have mentioned
already). Recall that by the same token the inclusive production rates
of \J and $\chi_c$ mesons become comparable to each other in spite
of the hierarchy of the wave functions ($S$-wave compared to $P$-wave).
As a consequence, although the two-gluon exchange $g{+}g\to\j{+}\j$ 
and one-gluon exchange $g{+}g\to\j{+}\j{+}g{+}g$ processes are of the 
same QCD order, their magnitudes are considerably different. 

%--------------------------
\section{Conclusions}
%--------------------------

We have considered the production of \J pairs at the LHC energies
via SPS and DPS processes taking into account several possible
contributing subprocesses.
We find it rather difficult to disentangle the SPS and DPS
modes on the basis of azimuthal or transverse momentum correlations:
the difference becomes only visible at sufficiently high $p_T$, where
the production rates are, indeed, very small.

Selecting large rapidity difference events looks more promising.
The leading order SPS contribution is localized inside the interval
$|\Delta{y}|\le 2$ (and continues to fall down steeply with increasing 
$|\Delta{y}|$), while the higher order contributions extending
beyond these limits are heavily suppressed by the color algebra and
do not constitute significant background for the DPS production.
%
%The two-gluon exchange contribution is very similar in shape to 
%the DPS contribution at large $\Delta y$, but is smaller by a factor
%of 20.
%The contribution from one-gluon exchange is even smaller by another
%order of magnitude and is practically negligible.

%We have shown that the differential distribution in $\Delta y$ for 
%the two-gluon exchange contribution is very similar in shape to that 
%for the DPS contribution at large $\Delta y$.
%The two-gluon exchange contribution,
%however, turned out to be much smaller in magnitude.
%In addition, we have shown that the contribution from one-gluon 
%exchange is practically negligible.

\acknowledgments
The authors thank Ivan Belyaev for useful discussions on many 
experimental issues.
This work is supported in part by the Polish
Grants DEC-2011/01/B/ST2/04535 and N N202 236 640,
by the Russian Foundation for Basic Research 
Grants No. 10-02-93118 and 11-02-01454,
by the FASI State contract 02.740.11.0244,
by the President of Russian Federation Grant No 3920.2012.2,
by the Ministry of Education and Sciences of Russian Federation
under agreement No. 8412,
and by the DESY Directorate in the framework of Moscow-DESY project
on Monte-Carlo implementations for HERA-LHC.

%\end{document}

\newpage

\begin{figure}\label{fusion}
\epsfig{figure=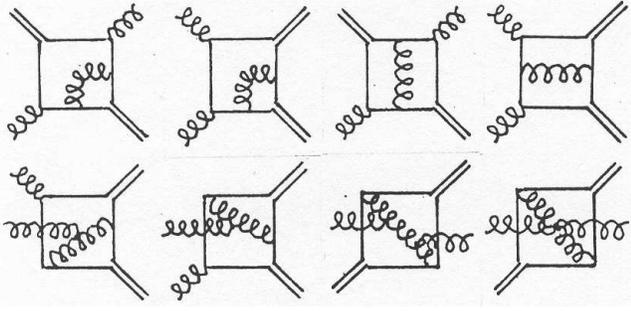,width=8.5cm}
\caption{Examples of Feynman diagrams representing the leading-order
 gluon-gluon fusion subprocess $gg\to\j\j$.}
\end{figure}

\begin{figure}\label{elastic}
\epsfig{figure=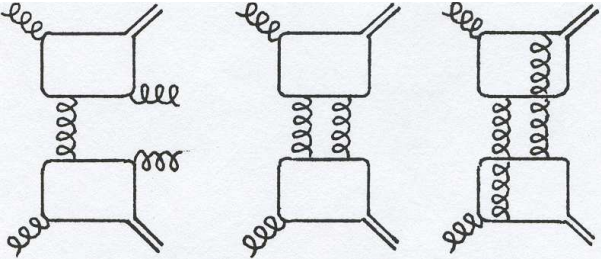,width=8.0cm}
\caption{Examples of Feynman diagrams representing the production 
of \J pairs in \pd ~gluon-gluon scattering via one-gluon exchange 
and two-gluon exchange mechanisms.}
\end{figure}  

\begin{figure}\label{ptmin}
\epsfig{figure=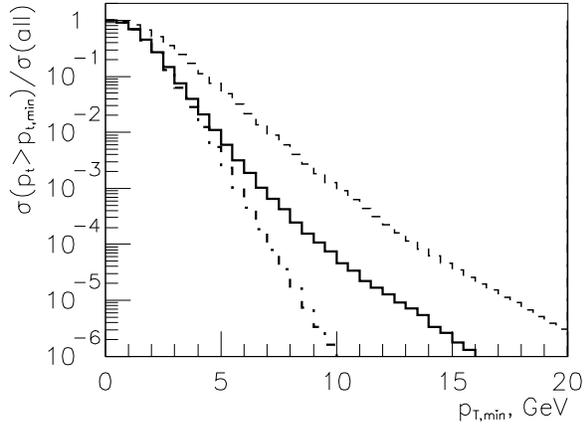,width=9.5cm}
\caption{Fraction of the production cross section left after 
imposing cuts on the \J transverse momentum.
Dashed curve, SPS mode under the requirement that one \J meson
has $p_T(\psi){>}p_{T,min}$;
dash-dotted curve, the square of the dashed curve;
solid curve, SPS mode under the requirement that both \Js have
$p_T(\psi){>}p_{T,min}$;
dotted curve, DPS mode under the requirement that both \Js have
$p_T(\psi){>}p_{T,min}$.}
\end{figure}

\begin{figure}\label{deltaphi}
\epsfig{figure=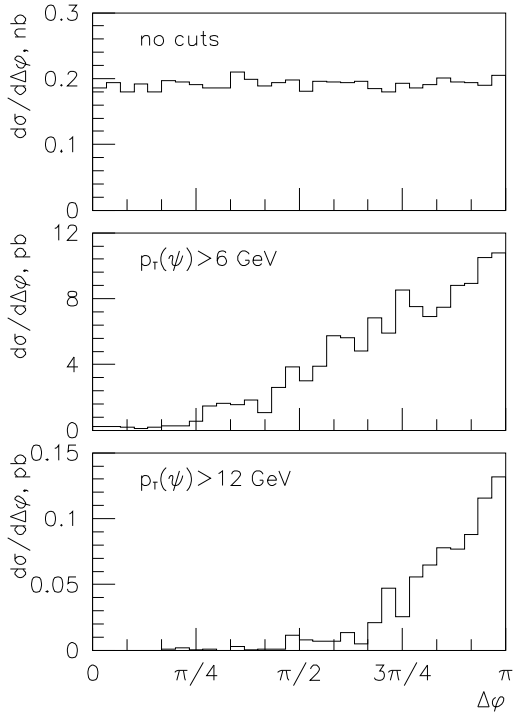,width=8.5cm}
\caption{Azimuthal angle difference distributions after imposing
cuts on the \J transverse momenta. Upper plot, no cuts;
middle plot, $p_{T,min}=6$ GeV; lower plot, $p_{T,min}=12$ GeV.}
\end{figure}

\begin{figure}\label{deltay}
\epsfig{figure=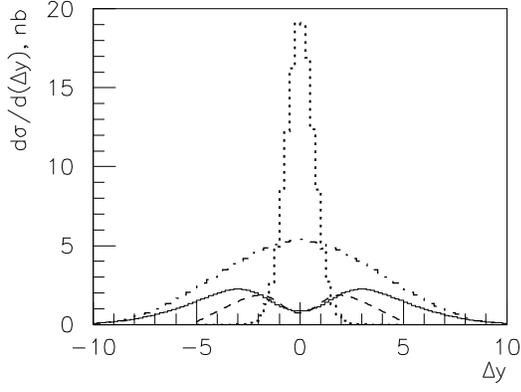,width=8.5cm}
\caption{Distribution over the rapidity difference between \J mesons.
Dotted curve, SPS 'box' contribution;
dashed curve, one-gluon exchange contribution multiplied by 1000;
solid curve, two-gluon exchange contribution multiplied by 25;
dash-dotted curve, DPS production.}
\end{figure}

\begin{figure}\label{y1vsy2}
\epsfig{figure=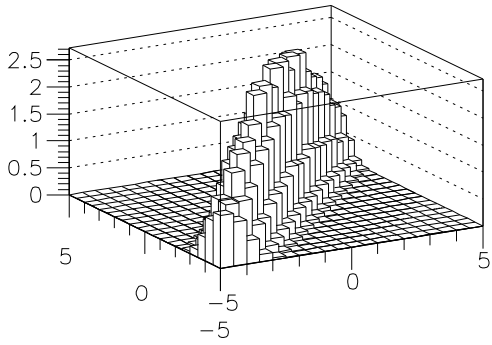,width=8.5cm}
\caption{Double differential distribution
$d\sigma/dy(\psi_1)dy(\psi_2)$ for the Leading-Order SPS 
production mode.}
\end{figure}

\end{document}